\documentclass[copyright,creativecommons]{eptcs}

\usepackage{iftex}

\usepackage{amsmath}
\usepackage{amssymb}
\usepackage{amsthm}
\usepackage{hyperref}
\usepackage{multicol}
\usepackage{lipsum}
\usepackage{wrapfig}
\usepackage[skip=5pt]{caption}
\usepackage{setspace}

\usepackage{ebproof} 
\usepackage{tikz}

\usepackage{algorithm}
\usepackage{algpseudocode}
\usepackage[strings]{underscore}
\usepackage[english]{babel}

\usepackage{sessiontypes}

\usetikzlibrary{shapes.geometric, arrows.meta}

\tikzstyle{rect} = [rectangle, rounded corners, minimum width=3cm, minimum height=1cm,text centered, draw=black, fill=black!10]

\theoremstyle{definition} \newtheorem{theorem}{Theorem}[section]
\theoremstyle{definition} \newtheorem{corollary}{Corollary}[theorem]
\theoremstyle{definition} \newtheorem{definition}[theorem]{Definition}
\theoremstyle{definition} \newtheorem{lemma}[theorem]{Lemma}
\theoremstyle{definition} \newtheorem{example}[theorem]{Example}
\theoremstyle{definition} \newtheorem{remark}[theorem]{Remark}

\newcommand{\nd}{\text{nd}}
\newcommand{\Sub}{\text{Sub}_\text{TD}}
\newcommand{\SubBU}{\text{Sub}}
\newcommand{\bigO}[1]{O\negmedspace\left(#1\right)}
\newcommand{\bigOmega}[1]{\Omega\negmedspace\left(#1\right)}
\newcommand{\unfold}[1]{\text{unfold}\negmedspace\left(#1\right)}
\newcommand{\cR}{\mathcal{R}}
\newcommand{\trans}[1]{\xrightarrow{#1}}

\newcommand{\od}{\text{od}}

\newcommand{\case}{\textbf{Case. }}

\newcommand{\kf}[1]{\ensuremath{\mathsf{#1}}}

\newcommand{\trinp}{\kf{?p}}
\newcommand{\trinc}{\kf{?c}}
\newcommand{\troutp}{\kf{!p}}
\newcommand{\troutc}{\kf{!c}}
\newcommand{\trbra}{\&}
\newcommand{\trsel}{\oplus}
\newcommand{\trend}{\kf{end}}

\newcommand{\RULE}[1]{[\textsc{#1}]}

\newcommand{\Tinter}{T^\kf{interface}}

\newcommand{\subt}{\leq_c}
\newcommand{\cC}{\mathcal{C}}

\newcommand{\etal}{\textit{et al.}}

\title{Three Subtyping Algorithms for Binary Session Types and their Complexity Analyses}
\author{Thien Udomsrirungruang
\institute{University of Oxford, Oxford, UK}
\email{thien.udomsrirungruang@keble.ox.ac.uk}
\and
Nobuko Yoshida
\institute{University of Oxford, Oxford, UK}
\email{nobuko.yoshida@cs.ox.ac.uk}
}

\def\titlerunning{Three Subtyping Algorithms for Binary Session Types and their Complexity Analyses}
\def\authorrunning{T. Udomsrirungruang \& N. Yoshida}

\begin{document}
\maketitle

\begin{abstract}
    Session types are a type discipline for describing and specifying communication behaviours of concurrent processes. \emph{Session subtyping}, firstly introduced by Gay and Hole \cite{Gay2005}, is widely used for enlarging typability of session programs. This paper gives the complexity analysis of three algorithms for subtyping of synchronous binary session types. First, we analyse the complexity of the algorithm from the original paper, which is based on an inductive tree search. We then introduce its optimised version, which improves the complexity, but is still exponential against the size of the two types. Finally, we propose a new quadratic algorithm based on a graph search using the concept of $\mathcal{XYZW}$-simulation, recently introduced by Silva \etal\ \cite{Silva2023}.
\end{abstract}

\section{Introduction} \label{section:introduction}
Session types \cite{Honda1998, Takeuchi1994} are a type discipline for describing and specifying communication behaviours of concurrent processes. They stem from the observation that many real-world processes communicate in a highly-structured manner; these communications may include sending and receiving messages, selection and branching from a set of labels, and recursion. These session types allow programs to be validated for type safety using typechecking algorithms.

With regards to the synchronous session type system, Gay and Hole \cite{Gay2005} proposed a subtyping scheme, and proved the soundness of the type system; Chen \etal\ \cite{Chen2017} later proved that the type system was complete (i.e. there is no larger subtyping relation between session types that is sound). Thus this notion of subtyping is of interest as it is most general. Our presentation of session types will use this system, as described in Gay and Hole's paper \cite{Gay2005}. For simplicity, we omit the \textit{channel type} $\chan{S}$: they can be adopted into the algorithms with little issue.

As an example, consider the following type, which represents an interface to a server which is able to respond to pings, as well as terminate itself:
\begin{align}
  \Tinter_1 &= \mu X. \bsel{
    \kf{respond}\!: \bin{\tend} X,
    \kf{exit}\!: \tend
  } \label{eq:tinter1}
\end{align}

\noindent Here, the channel offers two choices of label to the process using it: (1) choosing $\kf{respond}$ then receiving a message of unit type $\tend$, or (2) choosing $\kf{exit}$ and terminating the channel. After performing (1), the $\mu$-recursion means that the channel reverts to its original state.

Next, consider the scenario where we upgrade the interface to the server such that it can be replicated.
\begin{align}
  \Tinter_2 &= \mu X. \bsel{
    \kf{respond}\!: \bin{\tend} X,
    \kf{exit}\!: \tend,
    \kf{replicate}\!: \bin{X} X
  } \label{eq:tinter2}
\end{align}

\noindent The interface offers another choice: (3) choosing $\kf{replicate}$ and receiving a new channel with the same type $\Tinter_2$. We can immediately see that any program that expects a channel of type $\Tinter_1$ will also work on a channel of type $\Tinter_2$. This is because the functionality of the new channel type is identical to the old type, as long as the $\kf{replicate}$ label is never chosen. Thus, it would make sense to accept $\Tinter_2$ in type checking. It turns out that $\Tinter_2$ is a subtype of $\Tinter_1$ (denoted $\Tinter_2 \subt \Tinter_1$), which characterises this property.

This subtyping relation is easy to see with a \textit{syntactic} approach: the shapes of the types are almost identical, except for one additional branch in $\Tinter_2$; we could then argue inductively that the subtyping relation holds. However, for some types, this is not possible:
\begin{align}
  \Tinter_3 &= \mu Y. \bsel{
    \kf{respond}\!: \bin{\tend} Y,
    \kf{exit}\!: \tend,
    \kf{replicate}\!: \bin{\Tinter_1} Y
  } \label{eq:tinter3}
\end{align}

In essence, $\Tinter_3$ is an interface that can be replicated to get copies, but those copies cannot themselves be replicated. We have that $\Tinter_2 \subt \Tinter_3$. Syntactic subtyping does not work because the types do not have the same shapes (namely, the messages sent in the $\kf{replicate}$ branch are $X$ and $\Tinter_1$ respectively). The difficulty of checking subtyping comes from the ability to "branch off" into multiple subterms: in the above example, checking the subterm $\bin{\Tinter_1} Y$ would require us to check both $\Tinter_1$ and $Y$. In certain cases this can lead to exponential complexity in the standard inductive algorithm given in \cite{Gay2005} (see Example \ref{ex:exponential_example}).\\[1mm]
\textbf{Contributions.} We first analyse the algorithm given in Gay and Hole's original paper on session type subtyping \cite[\S 5.1]{Gay2005}, improving the bound given in \cite[\S 5.1]{Lange2016}. Second, we propose its optimised version, in the style of \cite[Fig. 21-4]{Pierce2002}. Both of these algorithms have worst-case exponential complexity, with the second algorithm having better complexity than the first. Third, we represent types as labeled transition systems, and formulate the subtyping problem as checking an $\mathcal{XYZW}$-simulation, similarly to \cite{Silva2023}; we then give a quadratic algorithm for subtyping by checking the validity of the simulation.

\section{Preliminaries}

We restate the definitions given in \cite{Gay2005}, which are used in this paper.

\begin{definition}[Session Types]
  Session types (denoted $S, S', T, U, V, W, \dots$) are defined by the following grammar:
  \begin{multicols}{3}
    \noindent
    \begin{alignat*}{2}
      S ::=&\: \tend&&\quad \text{(inaction)}\\
          |&\: \bin{S_1, \dots, S_n} S&&\quad \text{(input)}\\
          |&\: \bout{S_1, \dots, S_n} S&&\quad \text{(output)}\\
    \end{alignat*}
    \begin{alignat*}{2}
          |&\: \bsel{l_i: S_i}_{1 \leq i \leq n}&&\quad \text{(selection)}\\
          |&\: \bbra{l_i: S_i}_{1 \leq i \leq n}&&\quad \text{(branching)}\\
    \end{alignat*}
    \begin{alignat*}{2}
        |&\: \mu X. S&&\quad \text{(recursive type)}\\
        |&\: X&&\quad \text{(type variable)}
      \end{alignat*}
  \end{multicols}
  \noindent $X, X_i, Y, Z, \dots$ range over a countable set of type variables. We require that all terms are contractive, i.e. $\mu X_1. \mu X_2. \dots \mu X_n. X_1$ is not allowed as a subterm for any $n \geq 1$. As shorthand, we use $\bin{\tilde T} S$ instead of $\bin {T_1, \dots, T_n} S$ when there is no ambiguity on $n$, and similarly for $\bout{\tilde T} S$.
\end{definition}

\begin{definition}[Unfolding]\label{def:unfold} $\unfold{\mu X. T} = \unfold{T[\mu X. T / X]}$, and $\unfold{T} = T$ otherwise.

Note that as all terms are contractive, this is well-defined.
\end{definition}


\begin{definition}[Coinductive subtyping]\label{def:coinductive_subtyping} A relation $\cR$ is a subtyping relation if the following rules hold, for all $T \cR U$:
  
  \begin{itemize}
    \item If $\unfold{T} = \bin{\tilde T'} S_1$ then $\unfold{U} = \bin{\tilde U'} S_2$, and $\tilde T'\cR \tilde U'$ and $S_1 \cR S_2$.
    \item If $\unfold{T} = \bout{\tilde T'} S_1$ then $\unfold{U} = \bout{\tilde U'} S_2$, and $\tilde U' \cR \tilde T'$ and $S_1 \cR S_2$.
    \item If $\unfold{T} = \bbra{l_i: T_i}_{1 \leq i \leq m}$ then $\unfold{U} = \bbra{l_i: U_i}_{1 \leq i \leq n}$, and $m \leq n$ and $\forall i \in \{1, \dots, m\}.\:T_i \cR U_i$.
    \item If $\unfold{T} = \bsel{l_i: T_i}_{1 \leq i \leq m}$ then $\unfold{U} = \bsel{l_i: U_i}_{1 \leq i \leq n}$, and $n \leq m$ and $\forall i \in \{1, \dots, n\}.\:T_i \cR U_i$.
    \item If $\unfold{T} = \tend$ then $\unfold{U} = \tend$.
  \end{itemize}
  \noindent Subtyping $\subt$ is defined by $S \subt T$ if $(S, T) \in \cR$ in some type simulation $\cR$. It immediately follows that subtyping is the largest type simulation.
\end{definition}

\begin{definition}[Coinductive equality]\label{def:coinductive_equality} $T =_c T'$ if $T \leq_c T'$ and $T' \leq_c T$.
\end{definition}

\begin{example}[Interface]\label{ex:subtyping_relation} We can now formally prove the subtyping relations mentioned in Section \ref{section:introduction}. Let us define
   $\cR = \{
      (\Tinter_2, \Tinter_3), (\bin \tend \Tinter_2, \bin \tend \Tinter_3), (\tend, \tend), (\Tinter_2, \Tinter_1),\break
      (\bin {\Tinter_2} {\Tinter_2}, \bin {\Tinter_1} {\Tinter_3}), (\bin \tend {\Tinter_2}, \bin \tend {\Tinter_1})
    \}$.
  We verify that the rules in Definition \ref{def:coinductive_subtyping} hold. Thus $T \cR U$ implies $T \subt U$. In particular, $\Tinter_2 \subt \Tinter_3$ and $\Tinter_2 \subt \Tinter_1$.
\end{example}

To reason about the running time of the subtyping algorithms, we define the size of a session type.

\begin{definition}[Size] $|\tend| = |X| = 1$, $|\mu X. T| = |T| + 1$; $|\bout{T_1, \dots, T_n}U| = |\bin{T_1, \dots, T_n}U| = \sum_{i=1}^n |T_i| + |U| + 1$; $|\bsel{l_1: T_1, \dots, l_n: T_n}| = |\bbra{l_1: T_1, \dots, l_n: T_n}| = \sum_{i=1}^n |T_i| + 1$.
\end{definition}

We will also need a notion of subterms. The following function is defined in~\cite{Gay2005}.

\begin{definition}[Bottom-up subterms]\label{def:bottom_up_subterms} The set of \textit{bottom-up subterms} of $T$ is defined inductively as:\\
    $\SubBU(\tend) = \{\tend\}$;
    $\SubBU(X) = \{X\}$;
    $\SubBU(\mu X. T) = \{\mu X. T\} \cup \{S[\mu X.T / X] \mid S \in \SubBU(T)\}$;\\
    $\SubBU(\bout{T_1, \dots, T_n}S) = \{\bout{T_1, \dots, T_n}S\} \cup \SubBU(S) \cup \bigcup_{i=1}^n \SubBU(T_i)$;\\
    $\SubBU(\bin{T_1, \dots, T_n}S) = \{\bin{T_1, \dots, T_n}S\} \cup \SubBU(S) \cup \bigcup_{i=1}^n \SubBU(T_i)$;\\
    $\SubBU(\bsel{l_1: T_1, \dots, l_n: T_n}) = \{\bsel{l_1: T_1, \dots, l_n: T_n}S\} \cup \bigcup_{i=1}^n \SubBU(T_i)$; and\\
    $\SubBU(\bbra{l_1: T_1, \dots, l_n: T_n}) = \{\bbra{l_1: T_1, \dots, l_n: T_n}S\} \cup \bigcup_{i=1}^n \SubBU(T_i)$.

    \noindent Then we define $\SubBU(T, U) = \SubBU(T) \cup \SubBU(U)$.
\end{definition}

\noindent Due the use of unfolds (Definition~\ref{def:unfold}) in our proof rules, it will be easier to reason with a definition of subterms that use an operation similar to unfolding in the treatment of recursive types. Thus, we give the analogue of~\cite[Definition 21.9.1]{Pierce2002} for session types. The full definition is Appendix~\ref{def:top_down_subterms_full}.

\begin{definition}[Top-down subterms]\label{def:top_down_subterms} The set of \textit{top-down subterms} of $T$ is defined: $\Sub(\mu X. T) = \{\mu X. T\} \cup \Sub(T [\mu X.T / X])$, with all other rules from Definition \ref{def:bottom_up_subterms} with $\SubBU$ replaced by $\Sub$.
\end{definition}

\section{Inductive subtyping algorithms}

\paragraph{Number of subterms.} In this section we show that the number of top-down subterms of a binary session type is linear. Proving this directly by induction is difficult as the definition of top-down subterm of $\mu X. T$ relies on the definition of a potentially larger term $T[\mu X.T/X]$.

We adapt the proofs in~\cite[Chapter 21.9]{Pierce2002}, which dealt with $\mu$-types in the lambda calculus. We will use bottom-up subterms as a proxy for top-down subterms: we will show that all top-down subterms are bottom-up subterms, and there are linearly many bottom-up subterms. Note that in this section we assume all substitutions are capture-avoiding. This can be done by alpha-conversion without changing the size of the term.

\begin{lemma}\label{thm:number_of_bottom_up_subterms_is_linear} $|\SubBU(T)| \leq |T|$.
  \begin{proof}
    By induction on the structure of $T$.

    \case $T = \tend$, or $T = X$. Trivial.

    \case $T = \mu X. T'$. Assuming the inductive hypothesis, we have
      $|\SubBU(T)| = |\{T\} \cup \{S[\mu X.T / X] \mid S \in \SubBU(T)\}|
                  \leq |\{T\}| + |\SubBU(T)|
                  = 1 + |T'|
                  = |T|$.

    \case $T = \bsel{T_1, \dots, T_n}$, or $T = \bbra{T_1, \dots, T_n}$. Assuming the inductive hypothesis, we have
      $|\SubBU(T)| \leq |\{T\}| + \sum_{i=1}^n |\SubBU(T_i)|
                  = 1 + \sum_{i=1}^n |T_i|
                  = |T|$.

    \case $T = \bout{T_1, \dots, T_n}S$ or $T = \bin{T_1, \dots, T_n}S$. Similar to the above.
  \end{proof}
\end{lemma}

\begin{lemma}\label{thm:substitution_in_subterms} If $S \in \SubBU(T[Q/X])$ then $S = S'[Q/X]$ for some $S' \in \SubBU(T)$, or $S \in \SubBU(Q)$.
  \begin{proof}
    By induction on the structure of $T$.

    \case $T = \tend$. Then $S = \tend$, so take $S' = \tend$.

    \case $T = Y$. If $Y = X$, then $S \in \SubBU(Q)$. Otherwise $S = Y$, so take $S' = Y$.

    \case $T = \bout{T_1, \dots, T_n}U$. Then either:
      \begin{itemize}
        \item $S = T[Q/X]$. Then $S' = T$.
        \item $S \in \SubBU(U[Q/X])$. Then, by the inductive hypothesis, either $S \in \SubBU(Q)$, or $S = S'[Q/X]$ for some $S' \in \SubBU(U)$. In the latter case, $S' \in \SubBU(T)$ by definition.
        \item $S \in \SubBU(T_i[Q/X])$. Similar.
      \end{itemize}

    \case $T = \bin{T_1, \dots, T_n}U$, $T = \bsel{T_1, \dots, T_n}$ or $T = \bbra{T_1, \dots, T_n}$. Similar to the above case.

    \case $T = \mu Y. T'$. We have $S \in \SubBU(\mu Y. T'[Q/X])$. By definition, either:
      \begin{itemize}
        \item $S = \mu Y. T'[Q/X]$. Take $S' = T = \mu Y. T'$.
        \item $S = S_1[(\mu Y. T'[Q/X]) / Y]$ for some $S_1 \in \SubBU(T'[Q/X])$. Then by the inductive hypothesis, either: 
        \begin{itemize}
          \item $S_1 \in \SubBU(Q)$. Then because our substitutions are capture-avoiding, $Y \notin \text{fv}(Q)$, so\break $S_1[(\mu Y. T'[Q/X]) / Y] = S_1$. Therefore $S = S_1 \in \SubBU(Q)$.
          \item $S_1 = S_2[Q/X]$ for some $S_2 \in \SubBU(T')$.\\
          Then $S = S_2[Q/X][(\mu Y. T'[Q/X]) / Y] = S_2[\mu Y. T' / Y][Q/X]$. Take $S' = S_2[\mu Y. T' / Y]$. (By definition $S' \in \SubBU(T)$.)
        \end{itemize}
      \end{itemize}
  \end{proof}
\end{lemma}

\begin{lemma}\label{thm:top_down_subset_bottom_up} $\Sub(S) \subseteq \SubBU(S)$.
  \begin{proof}
    Similar to \cite[Prop. 21.9.10]{Pierce2002}.
    We need to show that each rule of $\Sub$ can be matched by $\SubBU$. All rules except for $\mu X.T$ are identical. For $\mu X.T$, we use Lemma~\ref{thm:substitution_in_subterms}: if $S \in \SubBU(T[\mu X.T/X])$ then either $S \in \SubBU(T)$, or $S = S'[\mu X.T/X]$ for some $S' \in \SubBU(T)$. The former is what we want; the latter is part of the rule for $\mu$ in $\SubBU$.
  \end{proof}
\end{lemma}

\begin{corollary}\label{thm:number_of_subterms_is_linear} $|\Sub(T)| \leq |T|$.
  Follows from Lemmas \ref{thm:top_down_subset_bottom_up} and \ref{thm:number_of_bottom_up_subterms_is_linear}.
\end{corollary}

\subsection{An inductive algorithm \cite{Gay2005}} \label{section:alg_1}

First, we introduce the algorithm for checking subtyping in the original paper by Gay and Hole \cite{Gay2005}. The paper introduces the \textit{algorithmic rules for subtyping} shown in Figure \ref{fig:subtyping_rules}. These rules prove judgements of the form $\Sigma \vdash T \leq U$, which is intuitively read: ``assuming the relations in $\Sigma$, we can deduce that $T$ is a subtype of $U$''. The paper then formalises this by proving soundness and completeness of the rules in Figure \ref{fig:subtyping_rules}, i.e. $T \subt U$ iff there is a proof tree deriving $\emptyset \vdash T \leq U$.

Thus, in the algorithm, the objective is to infer this rule. To do this, it builds the proof tree bottom-up, using rules in Figure \ref{fig:subtyping_rules} with \RULE{AS-Assump} used with highest priority, and ties between \RULE{AS-RecL} and \RULE{AS-RecR} broken arbitrarily (we will assume that \RULE{AS-RecL} takes priority). All other rules are applicable on disjoint sets of judgements, so there is no further ambiguity.

\begin{figure}
  \centering
  \small
  \setstretch{3}
  \begin{prooftree}
    \hypo{T \leq U \in \Sigma}
    \infer1[\RULE{AS-Assump}]{ \Sigma \vdash T \leq U }
  \end{prooftree}\hspace{2em}%
  \begin{prooftree}
    \infer0[\RULE{AS-End}]{ \Sigma \vdash \tend \leq \tend }
  \end{prooftree}

  \begin{prooftree}
    \hypo{\Sigma, \mu X.T \leq U \vdash T [\mu X.T/X] \leq U}
    \infer1[\RULE{AS-RecL}]{\Sigma \vdash \mu X.T \leq U}
  \end{prooftree}\hspace{2em}%
  \begin{prooftree}
    \hypo{\Sigma, T \leq \mu X.U \vdash T \leq U[\mu X. U/X]}
    \infer1[\RULE{AS-RecR}]{\Sigma \vdash T \leq \mu X.U}
  \end{prooftree}

  \begin{prooftree}
    \hypo{\Sigma \vdash \tilde T \leq \tilde U}
    \hypo{\Sigma \vdash V \leq W}
    \infer2[\RULE{AS-In}]{\Sigma \vdash \bin{\tilde T} V \leq \bin{\tilde U} W}
  \end{prooftree}\hspace{2em}%
  \begin{prooftree}
    \hypo{\Sigma \vdash \tilde U \leq \tilde T}
    \hypo{\Sigma \vdash V \leq W}
    \infer2[\RULE{AS-Out}]{\Sigma \vdash \bout{\tilde T} V \leq \bout{\tilde U} W}
  \end{prooftree}

  \begin{prooftree}
    \hypo{m \leq n}
    \hypo{\forall i \in {1, \dots, m}. \Sigma \vdash S_i \leq T_i}
    \infer2[\RULE{AS-Bra}]{\Sigma \vdash \bbra{l_i: S_i}_{1 \leq i \leq m} \leq \bbra{l_i: T_i}_{1 \leq i \leq n}}
  \end{prooftree}\hspace{2em}%
  \begin{prooftree}
    \hypo{m \leq n}
    \hypo{\forall i \in {1, \dots, m}. \Sigma \vdash S_i \leq T_i}
    \infer2[\RULE{AS-Sel}]{\Sigma \vdash \bsel{l_i: S_i}_{1 \leq i \leq n} \leq \bsel{l_i: T_i}_{1 \leq i \leq m}}
  \end{prooftree}
  \vspace{5pt}
  \caption{Algorithmic rules for subtyping, taken from \cite{Gay2005}.}
  \label{fig:subtyping_rules}
\end{figure}

Gay and Hole's proof of termination~\cite[Lemma 10]{Gay2005} contains the following fact, using $\SubBU$ instead of $\Sub$. However, with our definition of $\Sub$, the proof is clearer:

\begin{lemma}\label{thm:only_derive_subterms} If $\Gamma \vdash T' \leq U'$ is produced from $\emptyset \vdash T \leq U$, then $T' \in \Sub(T, U)$ and $U' \in \Sub(T, U)$, and for all $V \leq W \in \Gamma$, $V \in \Sub(T, U)$ and $W \in \Sub(T, U)$.

  \begin{proof}
    Verify this for each rule in Figure \ref{fig:subtyping_rules}. The result follows from transitivity of the subterm relation.
  \end{proof}
\end{lemma}

\begin{definition}[Nesting depth]
    $\nd(\tend) = \nd(X) = 1$;
    $\nd(\mu X. T) = \nd(T) + 1$;
    $\nd(\bin{T_1, \dots, T_n} U) = \nd(\bout{T_1, \dots, T_n} U) = \max (\{\nd(T_i) \mid 1 \leq i \leq n\} \cup \{\nd(U)\}) + 1$; and
    $\nd(\bsel{l_1: T_1, \dots, l_n: T_n}) = \nd(\bbra{l_1: T_1, \dots, l_n: T_n}) = \max (\{\nd(T_i) \mid 1 \leq i \leq n\}) + 1$.
\end{definition}

Using the notion of nesting depth, Gay and Hole proceed to prove termination, as follows. Observe that when generating the premise $\Gamma' \vdash V' \leq W'$ above $\Gamma \vdash V \leq W$, either
  $|\Gamma'| > |\Gamma|$; or
  $|\Gamma'| = |\Gamma|$ and $\nd(V') < \nd(V)$.

\noindent Thus pairs $(\Gamma, \nd(V))$ for judgements $\Gamma \vdash V \leq W$ are distinct along any path from the root to any leaf of the proof tree. Termination follows by observing that both $|\Gamma|$ and $\nd(V)$ are bounded. We may extend this to a complexity bound as follows.

\begin{theorem} The upper bound of the worst-case complexity of Gay and Hole's subtyping algorithm is $\bigO{n^{n^3}}$, where $n$ is the sum of the sizes of the two inputs.
  \begin{proof}
    When $\Gamma \vdash V \leq W$ is generated from the rule $\emptyset \vdash T \leq U$, we have:
    \begin{itemize}
      \item The number of possible judgements in $\Gamma$ is $|\Sub(T, U)|^2$ by Lemma \ref{thm:only_derive_subterms}.
      \item The number of possible values of $V$ is $|\Sub(T, U)|$, thus there are only $|\Sub(T, U)|$ possible values of $\nd(V)$.
    \end{itemize}
    \noindent Therefore the height of the tree is bounded by $(|T| + |U|)^3$, using Corollary \ref{thm:number_of_subterms_is_linear}, and the branching factor is $\bigO{|T| + |U|}$, so the upper bound of the worst-case complexity is $\bigO{n^{n^3}}$, taking $n = |T| + |U|$.
  \end{proof}
\end{theorem}

\subsection{An algorithm with memoization} \label{section:alg_2}

\begin{figure}
  \begin{scriptsize}
    \begin{minipage}[t]{.43\textwidth}
      \begin{algorithmic}[1]
        \Function{Subtype}{$\Delta, \Sigma, T, U$} \label{alg:memoized_subtyping_start}
          \If{$\Delta = \text{false}$}
            \State \Return false
          \EndIf
          \If{$\Sigma \vdash T \leq U \in \Delta$} \Comment{Memoization of inferences}
            \State \Return $\Delta$
          \EndIf
          \State $\Delta \gets \Delta \cup \Sigma \vdash T \leq U \in \Delta$ \Comment {Add to the memoized set} \label{alg:memoized_subtyping_memo_end}
          \If{$T \leq U \in \Sigma$}
            \State \Return $\Delta$
          \ElsIf{$T = \tend$ and $U = \tend$}
            \State \Return $\Delta$
          \ElsIf{$T = \mu X. T'$}
            \State \Return \Call{Subtype}{$\Delta, \Sigma \cup \{T \leq U\}, T'[T/X], U$}
          \ElsIf{$U = \mu X. U'$}
            \State \Return \Call{Subtype}{$\Delta, \Sigma \cup \{T \leq U\}, T, U'[U/X]$}
          \ElsIf{$T = \bin{\tilde T}V$ and $U = \bin{\tilde U}W$}
            \For{$(T_i, U_i) \gets (\tilde T, \tilde U)$}
              \State $\Delta \gets \Call{Subtype}{\Delta, \Sigma, T_i, U_i}$
            \EndFor
            \State \Return \Call{Subtype}{$\Delta, \Sigma, V, W$}
          \algstore{part1}
      \end{algorithmic}
    \end{minipage}
    \hspace{.02\textwidth}
    \begin{minipage}[t]{0.55\textwidth}
      \begin{algorithmic}[1]
          \algrestore{part1}
          \ElsIf{$T = \bout{\tilde T}V$ and $U = \bout{\tilde U}W$}
            \For{$(T_i, U_i) \gets (\tilde T, \tilde U)$}
              \State $\Delta \gets \Call{Subtype}{\Delta, \Sigma, U_i, T_i}$
            \EndFor
            \State \Return \Call{Subtype}{$\Delta, \Sigma, V, W$}
          \ElsIf{$T = \bbra{l_i: T_i}_{1\leq i\leq m}$ and $U = \bbra{l_i: U_i}_{1\leq i\leq n}$ and $m \leq n$}
            \For{$i \gets 1..m$}
              \State $\Delta \gets \Call{Subtype}{\Delta, \Sigma, T_i, U_i}$
            \EndFor
            \State \Return $\Delta$
          \ElsIf{$T = \bsel{l_i: T_i}_{1\leq i\leq n}$ and $U = \bsel{l_i: U_i}_{1\leq i\leq m}$ and $m \leq n$}
            \For{$i \gets 1..m$}
              \State $\Delta \gets \Call{Subtype}{\Delta, \Sigma, T_i, U_i}$
            \EndFor
            \State \Return $\Delta$
          \Else
            \State \Return false
          \EndIf
        \EndFunction
      \end{algorithmic}
    \end{minipage}
  \end{scriptsize}

  \caption{A memoized subtyping algorithm.}
  \label{fig:memoized_subtyping_algorithm}
\end{figure}

A way to optimise the first algorithm is to treat the proof tree like a proof DAG: as identical nodes will have the same subtrees, we can search for their proofs only once. The algorithm in Figure \ref{fig:memoized_subtyping_algorithm} performs a depth-first search of the proof tree, ignoring nodes that have been seen before. This is done by keeping a set $\Delta$ of visited nodes.

\begin{theorem} The algorithm in Fig. \ref{fig:memoized_subtyping_algorithm} has worst-case time complexity at most $2^{\bigO{n^2}}$.
  \begin{proof}
    Observe that the runtime is proportional to $|\Delta|$ at the end of the program. As an upper bound, by Lemma \ref{thm:only_derive_subterms}, there are $|\Sub(T, U)|^2$ possible judgements in $\Sigma$, and $|\Sub(T, U)|$ possible terms for $T$ and $U$, so $\Delta$ has size at most $2^{|\Sub(T, U)|^2} \cdot |\Sub(T, U)|^2 = 2^{\bigO{n^2}}$, again by Corollary \ref{thm:number_of_subterms_is_linear}.
  \end{proof}
\end{theorem}

\begin{example}\label{ex:exponential_example} To show for certain that the algorithms so far are exponential in complexity, we will show that the following construction takes exponential time for the two subtyping algorithms presented:
  \begin{alignat*}{3}
    & T_k \subt T_{k+1} \quad & \text{where} \quad && T_k &:= \mu X. \bin{\mu Y_{k-1}. V_{k-1}} \bin{\mu Y_{k-2}. V_{k-2}} \dots \bin{\mu Y_1. V_1} \bin{\mu Y_0. V_0} X\\
    & & \text{and} \quad && V_l &:= \underbrace{\bin{\mu Z. \bin{Z} Z}. \dots \bin{\mu Z. \bin{Z} Z}}_\textrm{$l$ times} X
  \end{alignat*}
\end{example}

We first show that the subtyping relation holds, by proving the stronger notion of coinductive equality.

\begin{lemma} In Example \ref{ex:exponential_example}, $T_k =_c \mu X. \bout{X} X$.
  \begin{proof}
    Let $U = \mu X. \bout{X} X$, and take $\cR = \{(T, U) \mid T \in \Sub(T_k)\}$ and $\cR' = \{(U, T) \mid T \in \Sub(T_k)\}$. It is easy to see that for all $T \in \Sub(T_k)$, $\unfold{T} = \bout{S_1} S_2$ for some $S_1, S_2 \in \Sub(T_k)$. Also we have $\unfold{U} = \bout{U} U$. Hence $(S_1, U), (S_2, U) \in \cR$ and $(U, S_1), (U, S_2) \in \cR'$. Thus $\cR$ and $\cR'$ are type simulations.
  \end{proof}
\end{lemma}

\begin{corollary} $T_k \subt T_{k+1}$.
  \begin{proof}
    By transitivity, $T_k =_c T_{k+1}$, and thus $T_k \subt T_{k+1}$ by Definition~\ref{def:coinductive_equality}.
  \end{proof}
\end{corollary}

Therefore, by completeness of the inductive rules \cite{Gay2005} it follows that the algorithm will construct a valid derivation of $\emptyset \vdash T_k \leq T_{k+1}$. As both algorithms presented so far will need to traverse every node in the tree at least once, we will now show that this proof tree has an exponential amount of nodes.

\begin{lemma}\label{lemma:exponential_nodes_in_proof_tree}
  Define:
    $W_r^k = V_r^k[T_k/X]$ and
    $S_r^k = \bin{\mu Y_{r-1}^k. W_{r-1}^k} \dots \bin{\mu Y_0^k. W_0^k} T_k$.
  Then for every sequence $\alpha_1, \dots, \alpha_l (0 \leq l < k)$ such that $0 \leq \alpha_i < k-i$:
  \begin{align}
\small    \left.
    \begin{aligned}
      &\{T_k \leq T_{k+1}, S_k^k \leq T_{k+1}\}
      \cup \{\mu Y_{\alpha_i}. W_{\alpha_i}^k \leq \mu Y_{\alpha_i+i}. W_{\alpha_i+i}^{k+1} \mid 1 \leq i \leq l\}\\
      \cup& \{W_{\alpha_i}^k \leq \mu Y_{\alpha_i+i}. W_{\alpha_i+i}^{k+1} \mid 1 \leq i \leq l\}
      \cup \{T_k \leq W_i^{k+1} \mid 1 \leq i \leq l\} \\
      \cup& \{S_{k-i}^k \leq T_{k+1} \mid 1 \leq i \leq l\}
    \end{aligned}
    \right\}
    &\vdash S_{k-l}^k \leq S_{k+1}^{k+1} \label{eq:exp_counterexample}
  \end{align}
  is derivable from $\emptyset \vdash T_k \leq T_{k+1}$ as the root.

  \begin{proof}
    By induction on $l$. For some $\alpha_1, \dots, \alpha_{k-1}$, let $\mathcal{S}_l$ be the set of inferences on the left side of (\ref{eq:exp_counterexample}).

    The base case $l=0$ is simple: $\mathcal{S}_0 = \{T_k \leq T_{k+1}, S_k^k \leq T_{k+1}\}$. The corresponding proof tree is\\[1mm]
    \makebox[\textwidth]{
      \centering
      \scriptsize
      \begin{prooftree}
        \hypo{\mathcal{S}_0 \vdash S_k^k \leq S_{k+1}^{k+1}}
        \infer1[\RULE{AS-RecR}]{T_k \leq T_{k+1} \vdash S_k^k \leq T_{k+1}}
        \infer1[\RULE{AS-RecL}]{\emptyset \vdash T_k \leq T_{k+1}}
      \end{prooftree}
    }\\

    For the inductive step, we will build the tree starting from $\mathcal{S}_{l-1} \vdash S_{k-(l-1)}^k \leq S_{k+1}^k$, for $l > 0$, using the inductive hypothesis. The following proof tree works, taking\\$\cC_1 = \mu Y_{\alpha_l}. W_{\alpha_l}^k \leq \mu Y_{\alpha_l+l}^{k+1}. W_{\alpha_l+l}^{k+1},\: \cC_2 = W_{\alpha_l}^k \leq \mu Y_{\alpha_l+l}^{k+1}. W_{\alpha_l+l}^{k+1},\: \cC_3 = T_k \leq W_l^{k+1},\: \cC_4 = S_{k-l}^k \leq T_{k+1}$:\\

\smallskip 

    \makebox[\textwidth]{
      \centering
      \scriptsize
      \begin{prooftree}
        \hypo{\dots}
        \hypo{\mathcal{S}_{l-1} \cup \{\cC_1, \cC_2, \cC_3, \cC_4\} \vdash S_{k-l}^k \leq S_{k+1}^{k+1}}
        \infer1[\RULE{AS-RecR}]{\mathcal{S}_{l-1} \cup \{\cC_1, \cC_2, \cC_3\} \vdash S_{k-l}^k \leq T_{k+1} = W_0^{k+1}}
        \hypo{\dots}
        \infer2[\RULE{AS-In}]{\vdots}
        \infer1[\RULE{AS-In}]{\mathcal{S}_{l-1} \cup \{\cC_1, \cC_2, \cC_3\} \vdash S_k^k \leq W_l^{k+1}}
        \infer1[\RULE{AS-RecL}]{\mathcal{S}_{l-1} \cup \{\cC_1, \cC_2\} \vdash W_0^k = T_k \leq W_l^{k+1}}
        \hypo{\dots}
        \infer2[\RULE{AS-In}]{\vdots}
        \infer1[\RULE{AS-In}]{\mathcal{S}_{l-1} \cup \{\cC_1, \cC_2\} \vdash W_{\alpha_l}^k \leq W_{\alpha_l + l}^{k+1}}
        \infer1[\RULE{AS-RecR}]{\mathcal{S}_{l-1} \cup \{\cC_1\} \vdash W_{\alpha_l}^k \leq \mu Y_{\alpha_l}^{k+1}. W_{\alpha_l+1}^{k+1}}
        \infer1[\RULE{AS-RecL}]{\mathcal{S}_{l-1} \vdash \mu Y_{\alpha_l}. W_{\alpha_l}^k \leq \mu Y_{\alpha_l+l}^{k+1}. W_{\alpha_l+l}^{k+1}}
        \infer2[\RULE{AS-In}]{\mathcal{S}_{l-1} \vdash S_{\alpha_l+1}^k \leq S_{\alpha_l+l+1}^{k+1}}
        \hypo{\dots}
        \infer2[\RULE{AS-In}]{\vdots}
        \infer1[\RULE{AS-In}]{\mathcal{S}_{l-1} \vdash S_{k-(l-1)}^k \leq S_{k+1}^{k+1}}
      \end{prooftree}
    }\\

    Observing that $\mathcal{S}_l = \mathcal{S}_{l-1} \cup \{\cC_1, \cC_2, \cC_3, \cC_4\}$ finishes the proof.
  \end{proof}
\end{lemma}

\begin{theorem} The lower bound of the worst-case complexity of both inductive algorithms in this section is $\Omega((\sqrt n)!)$.
\begin{proof}
  Consider Example \ref{ex:exponential_example}. The complexity is at least the number of distinct nodes in the proof tree. Lemma \ref{lemma:exponential_nodes_in_proof_tree} shows that there are $\Omega(k!)$ such nodes, by observing that each sequence $\alpha_1, \dots, \alpha_l$ yields a distinct set $\mathcal{S}_l$. As $|T_k| + |T_{k+1}| = \Theta(k^2)$, we conclude that both algorithms on inductive trees run in worst-case exponential time, i.e. $\Omega((\sqrt n)!)$.
\end{proof}
\end{theorem}

\section{Quadratic subtyping algorithm} \label{section:alg_3}

We exploit the coinductive nature of subtyping to yield a quadratic algorithm. Firstly, we translate the constructs from Definition \ref{def:coinductive_subtyping} into a labelled transition system (LTS), so that subtyping is defined as a simulation-like relation.

\begin{definition}\label{def:subtyping_lts} The type LTS is defined as in Figure \ref{fig:lts_rules}. For a type $T$, the type LTS for $T$ is the part of the LTS that is reachable from $T$. 
\end{definition}

\begin{figure}
  \centering
  \small
  \setstretch{3}
  \begin{prooftree}
    \hypo{\unfold{T} = \tend}
    \infer1[\RULE{G-E}]{T \trans{\trend} \kf{Skip}}
  \end{prooftree}\hspace{.4em}%
  \begin{prooftree}
    \hypo{\unfold{T} = \bin{T_1, \dots, T_n}U}
    \infer1[\RULE{G-IC}]{T \trans{\trinc} U}
  \end{prooftree}\hspace{.4em}%
  \begin{prooftree}
    \hypo{\unfold{T} = \bin{T_1, \dots, T_n}U}
    \hypo{1 \leq i \leq n}
    \infer2[\RULE{G-IP}]{T \trans{\trinp_i} T_i}
  \end{prooftree}

  \begin{prooftree}
    \hypo{\unfold{T} = \bout{T_1, \dots, T_n}U}
    \infer1[\RULE{G-OC}]{T \trans{\troutc} U}
  \end{prooftree}\hspace{.4em}%
  \begin{prooftree}
    \hypo{\unfold{T} = \bout{T_1, \dots, T_n}U}
    \hypo{1 \leq i \leq n}
    \infer2[\RULE{G-OP}]{T \trans{\troutp_i} T_i}
  \end{prooftree}

  \begin{prooftree}
    \hypo{\unfold{T} = \bbra{l_i: T_i}_{1 \leq i \leq m}}
    \hypo{1 \leq j \leq m}
    \infer2[\RULE{G-B}]{T \trans{\trbra l_j} T_j}
  \end{prooftree}\hspace{.4em}%
  \begin{prooftree}
    \hypo{\unfold{T} = \bsel{l_i: T_i}_{1 \leq i \leq m}}
    \hypo{1 \leq j \leq m}
    \infer2[\RULE{G-S}]{T \trans{\trsel l_j} T_j}
  \end{prooftree}
  \caption{Rules for the type LTS.}
  \label{fig:lts_rules}
\end{figure}

The above definition gives us a graphical representation of types, which will be easier to work with. We show that the size of this LTS is linear:

\begin{lemma} The number of nodes in the type LTS for $T$ is $O(|T|)$.
  \begin{proof}
    Note that $T \trans{\alpha} T'$ implies $T' \in \Sub(T)$ or $T' = \kf{Skip}$, thus all nodes reachable from $T$ are elements of $\Sub(T) \cup \{\kf{Skip}\}$. The result follows from Corollary \ref{thm:number_of_subterms_is_linear}.
  \end{proof}
\end{lemma}

\begin{lemma}\label{thm:number_of_edges_in_subtyping_lts_is_linear} The number of edges in the type LTS for $T$ is $O(|T|)$.
  \begin{proof}
    Define the following function, which is the out-degree of an unfolded type: $\od(X) = 0$; $\od(\tend) = 1$; $\od(\bout{T_1, \dots, T_n}U) = \od(\bin{T_1, \dots, T_n}U) = n+1$; and $\od(\bsel{l_1: T_1, \dots, l_n: T_n}) = \od(\bbra{l_1: T_1, \dots, l_n: T_n}) = n$. Also, $\od(\mu X. T) = 0$ as it is not an unfolded type.

    Then, the number of edges in the LTS is
    $\sum_{U \text{ reachable from } T} \od(\unfold{U}) \leq \sum_{U \in \Sub(T)} \od(U) \leq \sum_{U \in \SubBU(T)} \od(U)$.

    Let $f(T) = \sum_{U \in \SubBU(V)} \od(T)$.

    We prove that $f(T) \leq 2|T| - 1$, by structural induction on $T$.

    \case $T = \tend$, or $T = X$. Then $\SubBU(T) = \{T\}$, so $f(T) \leq 2|T| - 1$.

    \case $T = \mu X. T'$. Then $f(T) = \od(T) + f(T')$. By the inductive hypothesis, $f(T') \leq 2|T'| - 1$. By definition, $\od(T) = 0$, so $\sum_{U \in \SubBU(T)} \od(U) \leq 2|T'| - 1 \leq 2|T| - 1$.

    \case $T = \bout{T_1, \dots, T_n} W$, $T = \bin{T_1, \dots, T_n} W$. Then $f(T) = \od(T) + f(W) + \sum_{i=1}^n f(T_i)$. By the inductive hypothesis, $f(W) \leq 2|W| - 1$ and $f(T_i) \leq 2|T_i| - 1$. By definition, $\od(T) = n+1$, so $f(T) \leq n + 1 + 2|W| - 1 + \sum_{i=1}^n (2|T_i| - 1) = 2(1 + |W| + \sum_{i=1}^n |T_i|) - 2 = 2|T| - 2 \leq 2|T| - 1$.

    \case $T = \bsel{T_1, \dots, T_n}$, $T = \bbra{T_1, \dots, T_n}$. Then $f(T) = \od(T) + \sum_{i=1}^n f(T_i)$. By the inductive hypothesis, $f(T_i) \leq 2|T_i| - 1$. By definition, $\od(T) = n$, so $f(T) \leq n + \sum_{i=1}^n (2|T_i| - 1) = 2\sum_{i=1}^n |T_i| - 1 \leq 2|T| - 1$.
  \end{proof}
\end{lemma}

In line with the treatment of context-free session types in Silva \etal\ \cite{Silva2023}, we can then rewrite Definition \ref{def:coinductive_subtyping} in terms of this representation.
In the language of the paper, this is a $\mathcal{XYZW}$-simulation, with
$\mathcal{X} = \{\trinc, \troutc, \trinp_i, \trbra l, \tend\},
\mathcal{Y} = \{\trinp_i, \trsel l, \tend\},
\mathcal{Z} = \alpha \in \{\troutp_i\},
\mathcal{W} = \alpha \in \{\troutp_i\}$.
We will then demonstrate how to check this $\mathcal{XYZW}$-simulation relation in quadratic time on a type LTS.\@

\begin{definition}\label{def:coinductive_subtyping_by_lts} $\cR$ is a subtyping relation if, for all $T \cR U$:
  
\begin{itemize}
  \item If $T \trans{\alpha} T'$ then $U \trans{\alpha} U'$, and $T' \cR U'$, for $\alpha \in \{\trinc, \troutc, \trinp_i, \trbra l, \tend\}$.
  \item If $U \trans{\alpha} U'$ then $T \trans{\alpha} T'$, and $T' \cR U'$, for $\alpha \in \{\trinp_i, \trsel l, \tend\}$.
  \item If $T \trans{\alpha} T'$ then $U \trans{\alpha} U'$, and $U' \cR T'$, for $\alpha \in \{\troutp_i\}$.
  \item If $U \trans{\alpha} U'$ then $T \trans{\alpha} T'$, and $U' \cR T'$, for $\alpha \in \{\troutp_i\}$.
\end{itemize}

$S \subt T$ if $(S, T) \in \cR$ in some type simulation $\cR$.
\end{definition}

It follows that Definitions \ref{def:coinductive_subtyping} and \ref{def:coinductive_subtyping_by_lts} are equivalent.

\begin{definition}\label{def:inconsistent_node} Call a pair $(T, U)$ \textit{inconsistent} if at least one of the following hold: (1) $T \trans{\alpha} T'$ and $U \not\trans{\alpha}$, for some $\alpha \in \{\trinc, \troutc, \trinp_i, \trbra l, \tend\}$; (2) $U \trans{\alpha} U'$ and $T \not\trans{\alpha}$, for some $\alpha \in \{\trinp_i, \trsel l, \tend\}$; (3) $T \trans{\alpha} T'$ and $U \not\trans{\alpha}$, for some $\alpha \in \{\troutp_i\}$; or (4) $U \trans{\alpha} U'$ and $T \not\trans{\alpha}$, for some $\alpha \in \{\troutp_i\}$.
\end{definition}

The LTS presentation gives rise to the following algorithm for checking $T \leq_c U$, where we want to find a consistent relation $\cR$ containing $(T, U)$; this can be extended to an algorithm that checks for any $\mathcal{XYZW}$-simulation on finite structures.

\begin{theorem}\label{thm:quadratic_binary_subtyping} $T \leq_c U$ can be checked in $O(n^2)$ time (where $n = |T| + |U|$).
  \begin{proof}
    Our algorithm is as follows: construct a graph on nodes $(T', U') \in (\Sub(T, U) \cup \{\kf{Skip}\}) \times (\Sub(T, U) \cup \{\kf{Skip}\})$. For each node, check whether it is inconsistent (Definition \ref{def:inconsistent_node}). Then, add an edge $(V, W) \rightarrow (V', W')$ if $V \cR W$ directly implies $V' \cR W'$ under Definiton \ref{def:coinductive_subtyping_by_lts}. If any inconsistent nodes are reachable from $(T, U)$, then $T \not\subt U$, otherwise $T \subt U$.

    To show correctness, observe that any set of consistent vertices closed under reachability is a type simulation, directly from Definition \ref{def:coinductive_subtyping_by_lts}. The minimal such set containing $(T, U)$, if it exists, must be the set of reachable nodes from $(T, U)$. Thus we can check if any inconsistent nodes are contained in this set to solve the problem.

    Note that there are $O((|T|+|U|)^2)$ nodes and edges in the graph, by Lemmas \ref{thm:number_of_subterms_is_linear} and \ref{thm:number_of_edges_in_subtyping_lts_is_linear}. Thus we can check $T \leq_c U$ in $O(n^2)$ time with a simple reachability search.
  \end{proof}
\end{theorem}

\begin{corollary}\label{thm:quadratic_binary_subtyping_of_all_subterms} $T' \leq_c U'$ for all subterms $T', U' \in \Sub(T, U)$ can be checked in $O(n^2)$ time.
  \begin{proof}
By finding the set of nodes in the above graph for which no inconsistent node is reachable, which can be done with a graph search from the inconsistent nodes in time linear in the size of the graph. 
  \end{proof}
\end{corollary}
\begin{figure}
  \centering
  \small
  \tikzstyle{arrow} = [thick,-{Latex[length=2mm, width=1.5mm]},>=stealth]
  \tikzstyle{arrow2} = [thick,{Latex[length=2mm, width=1.5mm]}-{Latex[length=2mm, width=1.5mm]},>=stealth]
  \begin{tikzpicture}[>=Latex, node distance=1.5cm]
    \node (start) [rect] {($\Tinter_2, \Tinter_3$)};
    \node (n1) [rect, right of=start, xshift=4cm] {$(\bin \tend {\Tinter_2}, \bin \tend {\Tinter_3})$};
    \node (n2) [rect, below of=n1] {$(\tend, \tend)$};
    \node (n3) [rect, below of=start] {$(\bin {\Tinter_2} {\Tinter_2}, \bin {\Tinter_1} {\Tinter_3})$};
    \node (n4) [rect, below of=n3] {$(\Tinter_2, \Tinter_1)$};
    \node (n5) [rect, below of=n2] {$(\bin \tend {\Tinter_2}, \bin \tend {\Tinter_1})$};
    \node (n6) [rect, right of=n2, xshift=2.5cm] {$(\kf{Skip}, \kf{Skip})$};
    \draw[arrow2] (start) -- (n1);
    \draw[arrow] (n1) -- (n2);
    \draw[arrow] (start) -- (n2);
    \draw[arrow2] (start) -- (n3);
    \draw[arrow] (n3) -- (n4);
    \draw[arrow] (n4) -- (n2);
    \draw[arrow] (n5) -- (n2);
    \draw[arrow2] (n4) -- (n5);
    \draw[arrow] (n2)--(n6);
  \end{tikzpicture}
  \caption{Graph for $(\Tinter_2, \Tinter_3)$.}
  \label{fig:graph_example}
\end{figure}
The next examples use the above algorithm to decide subtyping for examples from Section \ref{section:introduction}.
\begin{example}
  The graph for $(\Tinter_2, \Tinter_3)$ is drawn in Figure \ref{fig:graph_example}. There are no inconsistent nodes, which shows that $\Tinter_2 \subt \Tinter_3$. Note the similarity to Example \ref{ex:subtyping_relation}.
\end{example}

\begin{example}
  Node $(\Tinter_1, \Tinter_2)$ is inconsistent because $\Tinter_2 \trans{\trsel \kf{replicate}} \bin{\Tinter_2} \Tinter_2$ but $\Tinter_1 \not\trans{\trsel \kf{replicate}}$ . Thus $\Tinter_1 \not\subt \Tinter_2$.
\end{example}


We could also represent this algorithm procedually, similarly to Figure \ref{fig:memoized_subtyping_algorithm}, differing only in what we keep track of to avoid repetition. In the inductive algorithm we store entire judgements of the form $\Gamma \vdash T \leq U$ (lines \ref{alg:memoized_subtyping_start}-\ref{alg:memoized_subtyping_memo_end}), but in this quadratic algorithm we only store visited nodes of the form $T \leq U$. The main difference is the number of possible values we could store: exponential in the former case and quadratic in the latter. Figure \ref{fig:quadratic_subtyping_algorithm} contains this presentation of the algorithm; here, we keep track of the visited nodes in a set $V$. Calling $\textsc{SubtypeQuad}(\emptyset, T, U)$ decides whether $T \subt U$. This procedural form is also very similar to the subtyping algorithm for $\mu$-lambda-terms in \cite[Fig. 21-4]{Pierce2002}.

However, the LTS presentation of the algorithm does have its benefits; by representing the types as a LTS (as in Theorem \ref{thm:quadratic_binary_subtyping}), we obtain a representation of types that eliminates the overhead of manipulating the types, to yield a truly quadratic algorithm.

\section{Related work}

The first algorithm for subtyping of
recursive function types dates back to Amadio and Cardelli \cite{Amadio1993}, where they give an exponential algorithm for recursive function type subtyping. Their algorithm inductively checks for $\alpha \rightarrow \beta \leq \gamma \rightarrow \delta$ by recursively checking that $\alpha \leq \gamma$ and $\beta \leq \delta$, unwrapping $\mu$-recursions, and keeping track of which pairs have been checked before.

Pierce \cite[Chapter 21.12]{Pierce2002} surveys the development of quadratic subtyping algorithms for these types. Notably, Kozen \etal\ \cite{Kozen1993} represent types as automata, then does a linear-time check on the product automaton of two types to check for subtyping, a method similar to our third algorithm (\S \ref{section:alg_3}).

The subtyping relation for synchronous session types was introduced by Gay and Hole \cite{Gay2005}, in which they showed that subtyping is sound and decidable. Their algorithm for subtyping (as presented here in \S \ref{section:alg_1}), is similar to Amadio and Cardelli's, adapted for session types. Later, Chen \etal \cite{Chen2017} proved that this relation is \textit{precise}: no strictly larger relation respects type safety.

Lange and Yoshida \cite{Lange2016} investigate subtyping for session
types by converting terms to a modal $\mu$-calculus formula which
represents its subtypes, then using a model checker to check for
subtyping. A short analysis of two other subtyping algorithms is also
provided, along with an empirical evaluation. The first is Gay and
Hole's algorithm \cite{Gay2005}, in which a doubly-exponential upper
bound is given, which we improve to a singly-exponential bound. The
second is an adaptation of Kozen's algorithm \cite{Kozen1993}, which
is similar to our LTS-based construction (\S \ref{section:alg_3}) in
that it builds a product automaton and checks for reachability. A
quadratic algorithm is given; however, the type system provided in
\cite{Lange2016} does not allow sending sessions as messages; instead,
sorts are used as payloads. Our algorithm is adapted to remove this
restriction. A comparison of complexity bounds in \cite{Lange2016} and
this paper is in Table \ref{fig:complexity_comparison}.

\begin{wraptable}{r}{0.5\textwidth}
  \small
  \centering
  \begin{tabular}{|c|c|c|c|}
    \hline
    Algorithm & \cite{Lange2016} & This paper \\
    \hline    
    Gay and Hole \cite{Gay2005} & $\bigO{n^{2^n}}$ & $\bigO{n^{n^3}}, \bigOmega{(\sqrt n)!}$ \\
    \hline
    \cite{Gay2005} with memoization & \textendash & $2^{\bigO{n^2}}, \bigOmega{(\sqrt n)!}$ \\
    \hline
    Coinductive subtyping & $\bigO{n^2}$ & $\bigO{n^2}$ \\
    \hline
  \end{tabular}

  \caption{Comparison between upper and lower bounds for the worst-case complexity given in \cite{Lange2016} and this paper to check the subtyping relation $T~\subt~U$ where $n = |T| + |U|$.}
  \label{fig:complexity_comparison}
\end{wraptable}
Subtyping for non-regular session types, in which there are infinitely many subterms, are explored by Silva \etal\ \cite{Silva2023}, in which the concept of $\mathcal{XYZW}$-simulation for session type subtyping is introduced; it is also used in this paper. A sound algorithm is introduced to semi-decide the subtyping problem for context-free session types, which is accompanied by an empirical evaluation. In general, subtyping for context-free session types is undecidable, as shown by Padovani \cite{Padovani2019}.

We believe that the results from this paper could be applied to other session typing schemes with little difficulty: for example, the subtyping relation for local synchronous multiparty session types \cite{Honda2008}, which is also sound and complete \cite{Ghilezan2019}.

\paragraph{Acknowledgements.} The authors thank 
PLACES'24 reviewers for their careful reading and comments, as well as
Diana Costa and Simon Gay for their valuable comments
on an earlier version of this paper.
The first author is supported by the Keble Association Grant. 
The second author is supported by 
EPSRC EP/T006544/2, EP/Y005244/1, EP/K011715/1, EP/K034413/1,
EP/L00058X/1, EP/N027833/2, EP/T014709/2, 
EP/V000462/1, EP/X015955/1 and  
Horizon EU TaRDIS 101093006.

\newpage
\bibliographystyle{eptcs}
\bibliography{SessionTypes}

\newpage

\appendix
\renewcommand\thefigure{\thesection.\arabic{figure}} 
\renewcommand\theHfigure{A\arabic{figure}}

\section{Appendix}
\setcounter{figure}{0}

\begin{definition}[Full version of Definition \ref{def:top_down_subterms}]\label{def:top_down_subterms_full} The set of \textit{top-down subterms} of $T$ is defined:
  \begin{align*}
    \Sub(\tend) &= \{\tend\}\\
    \Sub(X) &= \{X\}\\
    \Sub(\mu X. T) &= \{\mu X. T\} \cup \Sub(T [\mu X.T / X])\\
    \Sub(\bout{T_1, \dots, T_n}S) &= \{\bout{T_1, \dots, T_n}S\} \cup \Sub(S) \cup \bigcup_{i=1}^n \Sub(T_i)\\
    \Sub(\bin{T_1, \dots, T_n}S) &= \{\bin{T_1, \dots, T_n}S\} \cup \Sub(S) \cup \bigcup_{i=1}^n \Sub(T_i)\\
    \Sub(\bsel{T_1, \dots, T_n}) &= \{\bout{T_1, \dots, T_n}S\} \cup \bigcup_{i=1}^n \Sub(T_i)\\
    \Sub(\bbra{T_1, \dots, T_n}) &= \{\bin{T_1, \dots, T_n}S\} \cup \bigcup_{i=1}^n \Sub(T_i)
  \end{align*}
\end{definition}

\begin{figure}
  \begin{algorithmic}[1]
    \Function{SubtypeQuad}{$\Gamma, T, U$}
      \If{$\Gamma = \text{false}$}
        \State \Return false
      \EndIf
      \If{$T \leq U \in \Gamma$}
        \State \Return $\Gamma$
      \EndIf
      \State $\Gamma \gets \Gamma \cup \{T \leq U\}$ 
      \If{$T = \tend$ and $U = \tend$}
        \State \Return $\Gamma$
      \ElsIf{$\unfold{T} = \bin{\tilde T}V$ and $\unfold{U} = \bin{\tilde U}W$}
        \For{$(T_i, U_i) \gets (\tilde T, \tilde U)$}
          \State $\Gamma \gets \Call{SubtypeQuad}{\Gamma, T_i, U_i}$
        \EndFor
        \State \Return \Call{SubtypeQuad}{$\Gamma, V, W$}
      \ElsIf{$\unfold{T} = \bout{\tilde T}V$ and $\unfold{U} = \bout{\tilde U}W$}
        \For{$(T_i, U_i) \gets (\tilde T, \tilde U)$}
          \State $\Gamma \gets \Call{SubtypeQuad}{\Gamma, U_i, T_i}$
        \EndFor
        \State \Return \Call{SubtypeQuad}{$\Gamma, V, W$}
      \ElsIf{$\unfold{T} = \bbra{l_i: T_i}_{1\leq i\leq m}$ and $\unfold{U} = \bbra{l_i: U_i}_{1\leq i\leq n}$ and $m \leq n$}
        \For{$i \gets 1..m$}
          \State $\Gamma \gets \Call{SubtypeQuad}{\Gamma, T_i, U_i}$
        \EndFor
        \State \Return $\Gamma$
      \ElsIf{$\unfold{T} = \bsel{l_i: T_i}_{1\leq i\leq n}$ and $\unfold{U} = \bsel{l_i: U_i}_{1\leq i\leq m}$ and $m \leq n$}
        \For{$i \gets 1..m$}
          \State $\Gamma \gets \Call{SubtypeQuad}{\Gamma, T_i, U_i}$
        \EndFor
        \State \Return $\Gamma$
      \Else
        \State \Return false
      \EndIf
    \EndFunction
  \end{algorithmic}

  \caption{The algorithm in Theorem \ref{thm:quadratic_binary_subtyping} written in procedural form.}
  \label{fig:quadratic_subtyping_algorithm}
\end{figure}

\begin{remark}
  The above results hold for the subtyping relation $\subt$. However, if coinductive equality $=_c$ is required, from Definition \ref{def:subtyping_lts} it is a bisimulation and so can be checked in $O(n \log n)$ time using a bisimulation checking algorithm\cite{Paige1987}.
\end{remark}

\end{document}